\documentclass[twocolumn,aps,pra,preprintnumbers,amsmath,amssymb]{revtex4-1}
\usepackage{natbib}
\usepackage{graphicx}
\usepackage{bm}
\usepackage{gensymb}
\usepackage{color}
\begin{document}
\title{Decoherence in a $\mathcal{PT}$-symmetric qubit }
\author{Junaid Majeed Bhat$^1$ }
\email{junnimajeed@gmail.com}
\author{ Muzaffar Qadir Lone$^{2}$}
\email{lone.muzaffar@uok.edu.in}
 \author{Sanjoy Datta$^1$}
 \author{Ahmed Farouk$^3$}
 \address{$^1$ National Institute of Technology Rourkela, Orissa-769008.\\
 $^2$Department of Physics, University of Kashmir, Srinagar-190006.\\
 $^3$Faculty of Computer and Information Sciences, Mansoura University, Egypt. }


\begin{abstract} 
We investigate the decoherence in a 
  $\mathcal{PT}$-symmetric qubit coupled with a bosonic bath. Using cannonical transformations, we map the non-Hermitian 
Hamiltonian representing the $\mathcal{PT}$-symmetric qubit to a spin boson model.
Identifying the parameter $\alpha$ that demarcates  the hermiticity and non-hermiticity in the model, we show that
the qubit does not decohere at the transition from real eigen spectrum to complex eigen spectrum. 
Using  a general class of spectral densities, the strong suppression  of decoherence is observed  due to both 
 vaccum and thermal fluctuations of the bath, and initial correlations as we approach  the  transition point.

\end{abstract}
\maketitle

\section{Introduction}

 The fundamentals of quantum mechanics were 
thought of just as an  academic interest but ever since more and 
more non-hermitian systems became experimentally accessible \cite{bender,feng,Ref1} the notion changed.
In fact, recent experiments have shown that the hermiticity postulate of quantum mechanics 
may not as fundamental as thought \cite{Ref2,Ref3}. It was just mere convenience to say that every quantum 
system should be represented by Hermitian operators as they have real spectrum but the converse 
is not necessarily true, one could have real eigen values with non-hermitian operators as well.
One of the examples are $\mathcal{PT}$-symmetric Hamiltonians which have been realized in many different setups, 
such as optical \cite{Ref4,Ref5}, optomechanical \cite{Ref6}, or microcavity-based experiments \cite{Ref7}. In a nutshell,
one could define $\mathcal{PT}$-symmetric systems to be those which are invariant under joint time reversal $\mathcal{T}$ 
and parity  $\mathcal{P}$ operations. It has been shown that $\mathcal{PT}$-symmetric Hamiltonians not only admit real spectrum 
but can also be mapped into hermitian Hamiltonians with suitable transformations \cite{Ref8}. 

When a quantum system of interest interacts with an
environment, its evolution becomes non-unitary and displays  decoherence \cite{HP}.
Decoherence is  the fundamental mechanism by which fragile superpositions are destroyed
thereby producing a quantum to classical transition \cite{sol,zurek2}. 
In fact, decoherence is one of the main obstacles for the preparation, observation, and
implementation of multi-qubit entangled states.
The intensive work on quantum information and computing in recent years has 
tremendously increased the
interest in exploring and controlling decoherence effects   
\cite{nat1,milb2,QA,diehl,verst,weimer,bellmo,FR2,FR3,FR4,FR5}.
A natural question would pertain to decoherence in $\mathcal{PT}$-symmetric systems and how  decoherence
varies with the change in ``amount of hermiticity" of the Hamiltonian.

It has been observed that non hermiticity leads to  slowing of decoherence \cite{Ref8,Refpt2} in the long time limit  of dynamics.
In this paper, for the first time we address the question pertaining decoherence in 
 $\mathcal{PT}$-symmetric qubit  without any approximation on dynamics. 
 We consider both the situations where qubit and bath are initially uncorrelated as well as correlated; we show that 
  the decoherence imparted by the initial correlations (as well as in uncorrelated case) is  significantly suppressed
  as we change the hermiticity in the model.

This paper is organised as follows. We introduce the $\mathcal{PT}$-symmetric qubit model in section \ref{1}.
This Hamiltonian depends on a parameter $\alpha$ which separates the real and complex eigen spectrum of the system.
We map the non-Hermitian Hamiltonian to spin boson model with suitable canonical transformations.  Assuming the system and bath at thermal equilibrium 
at time before $t=0$, we make a projective measurement on the system only, which result in a  bath state that 
depends on state vector of the system. In section 
\ref{2}, we study the 
decoherence due to this state dependent bath as well as due to uncorrelated initial states, and 
show that decoherence due to these initial correlations are strongly modified
by the change in the parameter $\alpha$ controlling the nature of the of the model Hamiltonian.
We finally conclude in section \ref{3}.
  
\section{$\mathcal{PT}$-symmetric Model Hamiltonian coupled with a bosonic bath}
\label{1}
  The system under consideration is a $\mathcal{PT}$-symmetric qubit coupled to a bosonic bath described as
\begin{eqnarray}
 H=H_S\otimes I_B +I_S\otimes H_B+H_I
\end{eqnarray}
where $H_s= i\alpha \sigma^z + \sigma^x $ is a $\mathcal{PT}$-symmetric qubit Hamiltonian \cite{Ref8,pati}. 
We see that $H_s$ has two eigenvalues $E_{\pm} = \pm\sqrt{1-\alpha^2}$. Thus, for $|\alpha|\leq 1$, we 
 will have real eigenvalues. $\alpha=1$ would therfore correspond to the transition point separating the real 
 and complex eigen spectrum.  For future references  $\alpha$ will be called hermiticity or hermiticity parameter
 and hence defines the hermiticity  in the Hamiltonian. $H_B=\sum\limits_k \omega_k b_k^\dagger b_k $ represents the bosonic bath with $b_k$ as an anhiliation operator of $k$th bosonic 
mode with energy $\omega_k$. The interaction between the qubit and bath is given by
$H_I=\sum\limits_k(i\alpha \sigma^z + \sigma^x)(g_k b_k+ g_k^{*} b_k^\dagger) $. 

This hamiltonian can be mapped to a hermitian hamiltonian,$H$ via an operator $T$ which preservers quantum canonical relations. Identifying, 
\begin{eqnarray}
T=\Delta^{\dagger}  \begin{pmatrix} s_+ & 0 \\0 & s_-\end{pmatrix} \Delta ,
\end{eqnarray}
with $s_{\pm}=\sqrt{2(1\pm\alpha)}$ and 
$   \Delta=\frac{1}{\sqrt{2}}\begin{pmatrix} i & 1 \\ -i & 1 \end{pmatrix} $ we can write,
\begin{eqnarray}
 \tilde{H}&=&TH_ST^{-1}\otimes I_B+I_S\otimes TH_BT^{-1}+TH_IT^{-1} \nonumber\\
  &=& E\sigma_x +\sum\limits_k \omega_k b_k^\dagger b_k +\sum\limits_k E \sigma_x(g_k b_k+ g_k^{*} b_k^\dagger)
\end{eqnarray}
with $E= \sqrt{1-\alpha^2}$. It is clear that the transformed Hamiltonian is Hermitian with $g_kE$ as the effective coupling.
Making the transformation $\sigma_x\rightarrow \sigma_z$, we get 
\begin{eqnarray}
 \tilde{H}=E\sigma_z +\sum\limits_k \omega_k b_k^\dagger b_k +\sum\limits_k E \sigma_z(g_k b_k+ g_k^{*} b_k^\dagger)
 \label{sp}
\end{eqnarray}
which is the  well known spin-boson model.  
  The distinctive feature of this dephasing model 
is that the average populations of the qubit states do not
depend on time. 

\section{Decoherence}
\label{2}
\subsection{Uncorrelated State}

Suppose at time $t=0$ the state of the composite system 
is described by the initial density matrix $\rho(0)$, then at time t the density matrix in interaction picture is given by
\begin{eqnarray}
 \rho(t) = U(t)\rho(0)U(t)^{\dagger}
\end{eqnarray}
where $U(t)=T~e^{-i\int_0^t dt^{\prime} H_I(t^{\prime})}$ is the time evolution operator,
$H_I(t)$ is the interaction Hamiltonian in the interaction picture and $T$ is the  chronological time ordering operator.  
Our main interest is to calculate the reduced density matrix of the system
by tracing over degrees of freedom of the bath:
\begin{eqnarray}
 \rho_s(t)= Tr_B[  U(t)\rho(0)U(t)^{\dagger} ]
\end{eqnarray}
We assume the  initial density matrix of the total system as a direct product state:
   \begin{equation}
    \varrho(0)=\varrho^{}_{S}(0)\otimes \varrho^{}_{B},
     \qquad
     \left. \varrho^{}_{B}=\text{e}^{-\beta H^{}_{B}}\right/Z^{}_{B}\, ,
     \label{Init-fact}
   \end{equation}
where $\beta=1/k^{}_{\text{B}}T$, and $Z^{}_{B}$ is the bath partition
function. Note that $\varrho^{}_{S}(0)$ may be a pure state as well as a
mixed state of the qubit.

Next  we write 
$U(t) = T e^{-i\int_0^{t} d\tau H_I(\tau)} = e^{i\phi(t)}e^{\sigma^z \hat{\Lambda}(t)}$ where $\phi(t)$
is a function of time only and $\hat{\Lambda}(t) = \sum_k[ \alpha_k(t)b_k - \alpha_k^*(t) b_k^{\dagger} ]$ with
$\alpha_k(t) = Eg_k \frac{e^{-i\omega_k t}-1}{\omega_k}$. Therefore, we can write for 
the qubit state  $|\psi\rangle= a |0\rangle + b |1\rangle$:

\begin{eqnarray}
 \rho_s(t) &=& {\rm Tr}_B[U(t) \rho(0) U(t)^{\dagger}] \nonumber \\
 &=& {\rm Tr}_B[ e^{\sigma^z \hat{\Lambda}(t)  } |\psi\rangle \langle \psi| \otimes 
 \varrho_B  e^{-\sigma^z \hat{\Lambda}(t)}   ] \nonumber \\
 &=& \begin{pmatrix}
|a|^2 & ab^* e^{-\gamma_1(t)}\\
ba^*e^{-\gamma_1(t)} & |b|^2
\end{pmatrix}
\end{eqnarray}
 with  $\gamma_1(t)$ defined as
  \begin{eqnarray}
  \gamma_1(t)=
    - \sum_{k}\ln \left\langle
       \exp\left[
       \alpha^{}_{k}(t) b^{\dagger}_{k} -
           \alpha^{*}_{k}(t) b^{}_{k}
           \right]\right\rangle^{}_{B},
   \label{Gamma(t)}
  \end{eqnarray}
 where the symbol $\langle \ldots \rangle^{}_{B}$ denotes averages taken
  with the bath
distribution $\varrho^{}_{B}$. After straightforward algebra  one finds
   \begin{equation}
 \gamma_1(t)= (1-\alpha^2) \int^{\infty}_{0} d\omega\,
  J(\omega)\,\coth\left(\beta\omega/2\right)\,
  \frac{1-\cos\omega t}{\omega^{2}}\, ,
      \label{Gamma(t)-contin}
   \end{equation}
 where the continuum limit of the bath modes is performed, and the spectral
 density $J(\omega)$ is introduced by the rule \cite{HP}
   $
    \sum_{k} 4|g^{}_{k}|^{2}\,f(\omega^{}_{k})=
     \int^{\infty}_{0} d\omega\, J(\omega)f(\omega).
     \label{J(omega)}
   $
Expression (\ref{Gamma(t)-contin}) is the exact result for the
 decoherence function in the
model (\ref{sp}) under the uncorrelated  initial condition
(\ref{Init-fact}). We thus observe that the decoherence function is scaled by the factor $E^2=1-\alpha^2$. Thus changing
$\alpha$ from zero to 1 results in the suppression of decoherence. At transition point from
hermitian to non-hermitian Hamiltonian $\alpha=\pm1$, no decoherence results making qubit state maximally robust. In the next 
section we see the same effect in correlated initial states.
  
\subsection{Correlated initial States  }

 \begin{figure*}[htb] 
 \begin{center}
{\includegraphics[width=2.2in,height=1.75in]{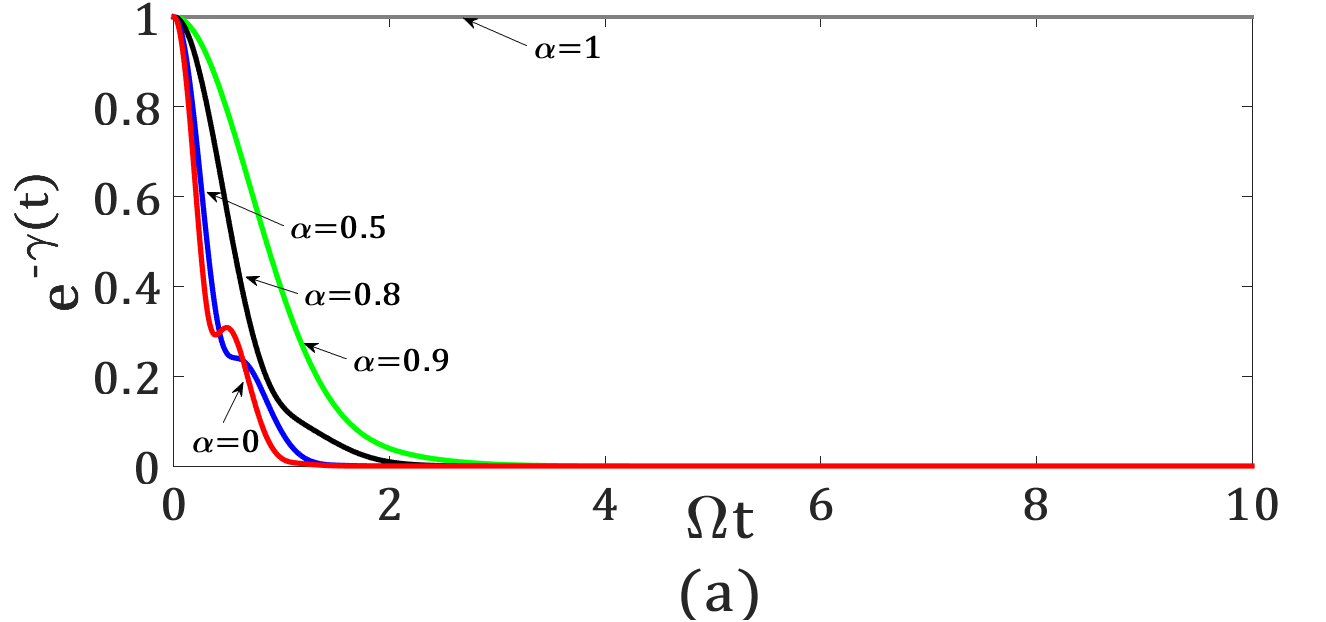}}\hspace*{10pt}
{\includegraphics[width=2.2in,height=1.75in]{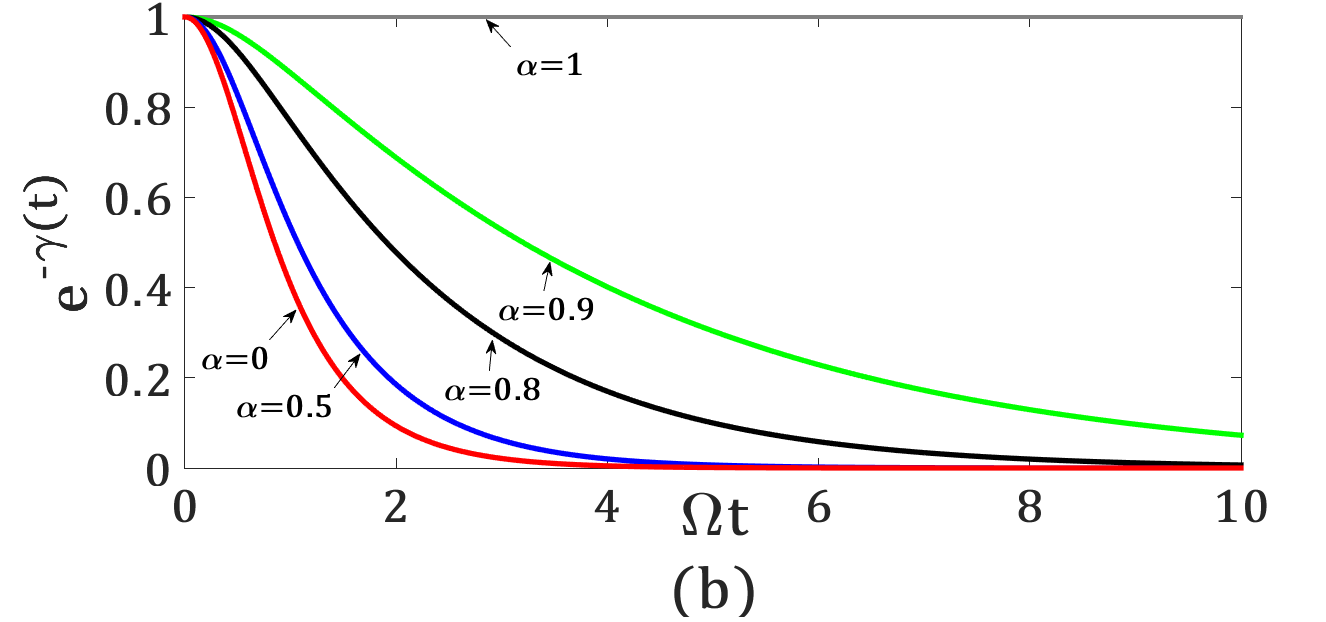}}\hspace*{10pt}
{\includegraphics[width=2.2in,height=1.75in]{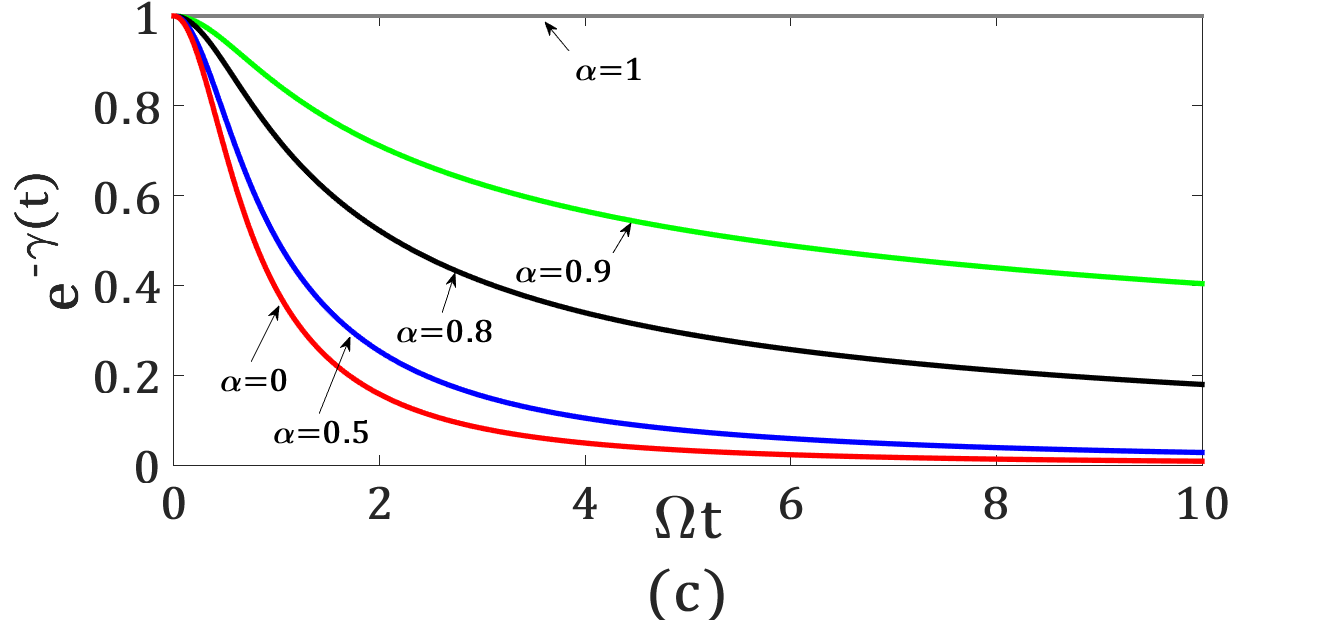}}
\end{center}
 \caption{Time dependence of exponential of the total decoherence function $e^{-\gamma(t)}$ for initially correlated state for different
 values of $\alpha$  
 with $\beta\omega_0=1$ and the initial condition $\langle\sigma_z\rangle=0, \Omega\beta=1$
 in all the  (a) subohmic  s=0.2 (b) ohmic  s=1 and (c) superohmic  s=2 cases. }
 \label{Fig1}
\end{figure*}

Next, we assume the total system plus bath are in thermal equilibrium state at time 
$t<0$, and a measurement is made on such state at time $t=0$, such that we have at $t=0$ \cite{Nielsen,muz,mozorov}

\begin{eqnarray}
 \rho(0)= \frac{1}{Z}\sum_m \Omega_m e^{-\beta H} \Omega_m^{\dagger} 
\end{eqnarray}
where $\Omega_m$ are the  projection operators on a desired state of system and/or  bath. 
$Z$ is the normalization constant called
as partition function. Now we make  particular case of selective measurement, 
a projection by taking \cite{muz}

\begin{eqnarray}
 \Omega_m = |\psi\rangle \langle \psi| \otimes  I_B 
\end{eqnarray}
where $I_B$ is the identity operation on the  state  of  bath and $|\psi\rangle$ is a pure state of the qubit. 
Therefore we write
\begin{eqnarray}
 \rho(0)= |\psi\rangle \langle \psi| \otimes \rho_B(\psi)
\end{eqnarray}
where $ \rho_B(\psi)= \frac{1}{Z}  \langle \psi | e^{-\beta H} |  \psi \rangle $ represents the density matrix 
of the bath and clearly depends on the state of the qubit $|\psi\rangle$.

\begin{eqnarray}
 \rho_s(t) &=& {\rm Tr}_B[U(t) \rho(0) U(t)^{\dagger}]\\
 &=& {\rm Tr}_B[ e^{\sigma^z \hat{\Lambda}(t)  } |\psi\rangle \langle \psi| \otimes 
 \rho_B(\psi)  e^{-\sigma^z \hat{\Lambda}(t)}   ]
\end{eqnarray}
Now to evaulate the above expression, we take 
the general state of the qubit as $|\psi\rangle= a |0\rangle + b |1\rangle$, while we relegate the derivation to appendix
\ref{apen}, 
we write

\begin{eqnarray}
 \rho_s(t) = \begin{pmatrix}
|a|^2 & ab^* F(t)\\
ba^* F^*(t) & |b|^2
\end{pmatrix}
\label{reduced}
\end{eqnarray}
\begin{widetext}
where 
\begin{eqnarray}
\!\!\! F(t) &=& [\frac{|a|^2 e^{-\beta \omega_0/2} e^{ i(1-\alpha^2) \Phi(t)} +  
|b|^2 e^{\beta \omega_0/2} e^{ -i (1-\alpha^2)\Phi(t)}  }
 {|a|^2 e^{-\beta \omega_0/2}+ |b|^2 e^{\beta \omega_0/2}} ]
  e^{-\gamma_1(t)}   \nonumber \\
\end{eqnarray}
with  $\Phi(t)=\sum\limits_{k}\frac{4|g_k|^2}{\omega_k^2}\sin\omega_k t $.
Using the fact $|a|^2+|b|^2=1$ and $\langle \sigma_z\rangle =|a|^2-|b|^2$, we can write 
\begin{eqnarray}
 \frac{|a|^2 e^{-\beta \omega_0/2} e^{ i (1-\alpha^2)\Phi(t)} +  |b|^2 e^{\beta \omega_0/2} e^{ -i (1-\alpha^2)\Phi(t)}  }
 {|a|^2 e^{-\beta \omega_0/2}+ |b|^2 e^{\beta \omega_0/2}}
 =\cos(1-\alpha^2)\Phi(t) - i \frac{ \sinh\frac{\beta \omega_0}{2} -\langle\sigma_z\rangle \cosh\frac{\beta \omega_0}{2}}
 {\cosh\frac{\beta \omega_0}{2}-\langle\sigma_z\rangle\sinh\frac{\beta \omega_0}{2}} \sin(1-\alpha^2)\Phi(t).
\end{eqnarray}

In order to get dephasing or time dependent frequency shift explicity we define

\begin{equation}
\tan[\chi(t)]= \frac{\sinh(\beta \omega_0/2) -
\langle\sigma_z\rangle \cosh(\beta \omega_0/2)}{\cosh(\beta \omega_0/2)-
\langle\sigma_z\rangle \sinh(\beta \omega_0/2)}\tan[(1-\alpha^2)\Phi (t)]
\end{equation}

so that $F(t)$ simplifies to $F(t) = e^{i\chi(t)} e^{-\gamma_1(t)-\gamma_{c}(t)} =e^{i\chi(t) - \gamma(t)} $
with $\gamma(t) = \gamma_1(t)+\gamma_{c}(t) $ and

\begin{eqnarray}
 \gamma_c(t) =-\frac{1}{2} \ln \Bigg[ 1-\frac{(1-\langle\sigma_z\rangle^2)\sin^2[(1-\alpha^2)\Phi (t)]} {\Big(\cosh(\beta \omega_0/2)
-\langle\sigma_z\rangle \sinh(\beta \omega_0/2)\Big)^2} \Bigg] 
\end{eqnarray}
\end{widetext}
The term $\gamma_1(t)$ represents the decoherence due to vaccum and thermal fluctuations of the bath while $\gamma_c(t)$
represents the decoherence due to initial correlations of the composite system. Thus we see that the
decoherence function $\gamma_1(t)$ gets scaled by factor of $1-\alpha^2$ {while as a different functional dependence on $\alpha$ is found for $\gamma_c(t)$}.The reduced dynamics of
$\mathcal{PT}$-symmetric qubit can be calculated as $T^{-1}\rho_s (t) T$.

  \begin{figure*}[htb] 
 \begin{center}
{\includegraphics[width=2.2in,height=1.75in]{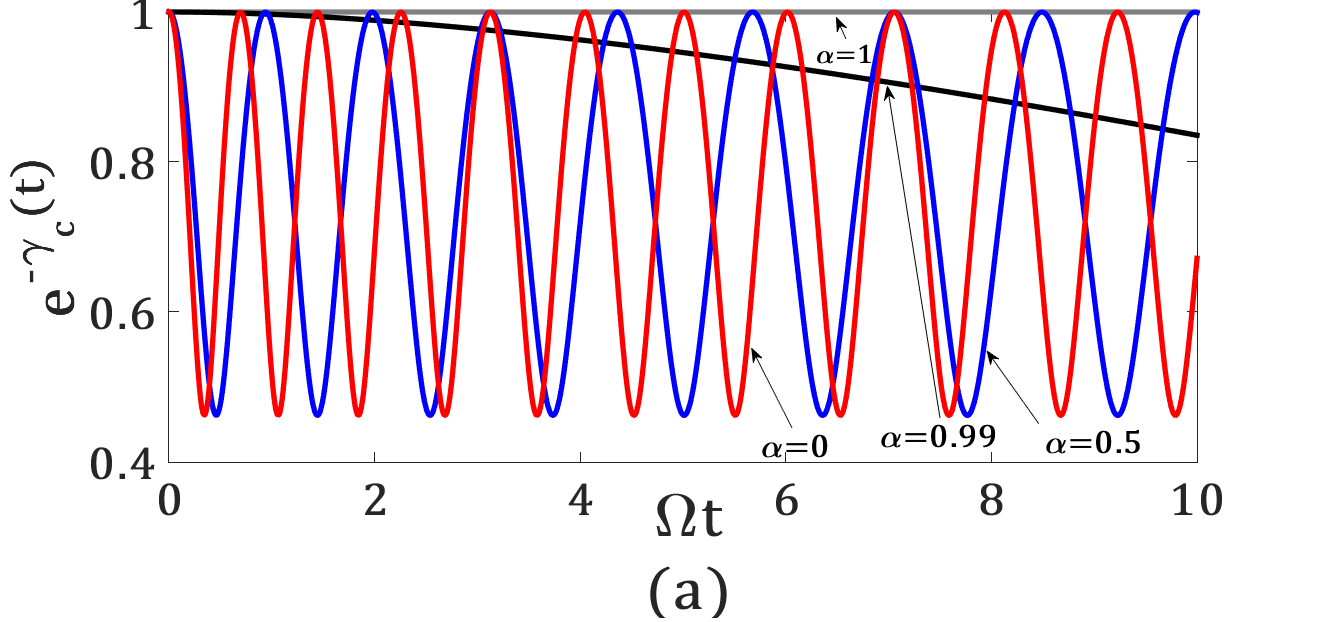}}\hspace*{10pt}
{\includegraphics[width=2.2in,height=1.75in]{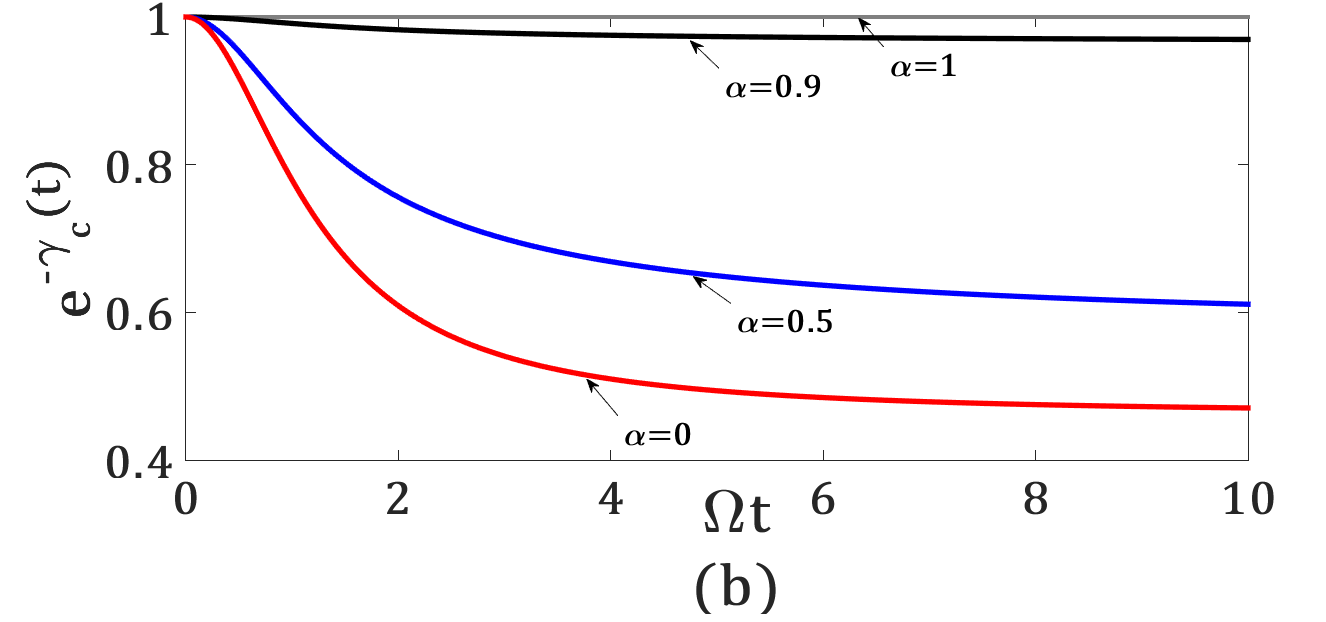}}\hspace*{10pt}
{\includegraphics[width=2.2in,height=1.75in]{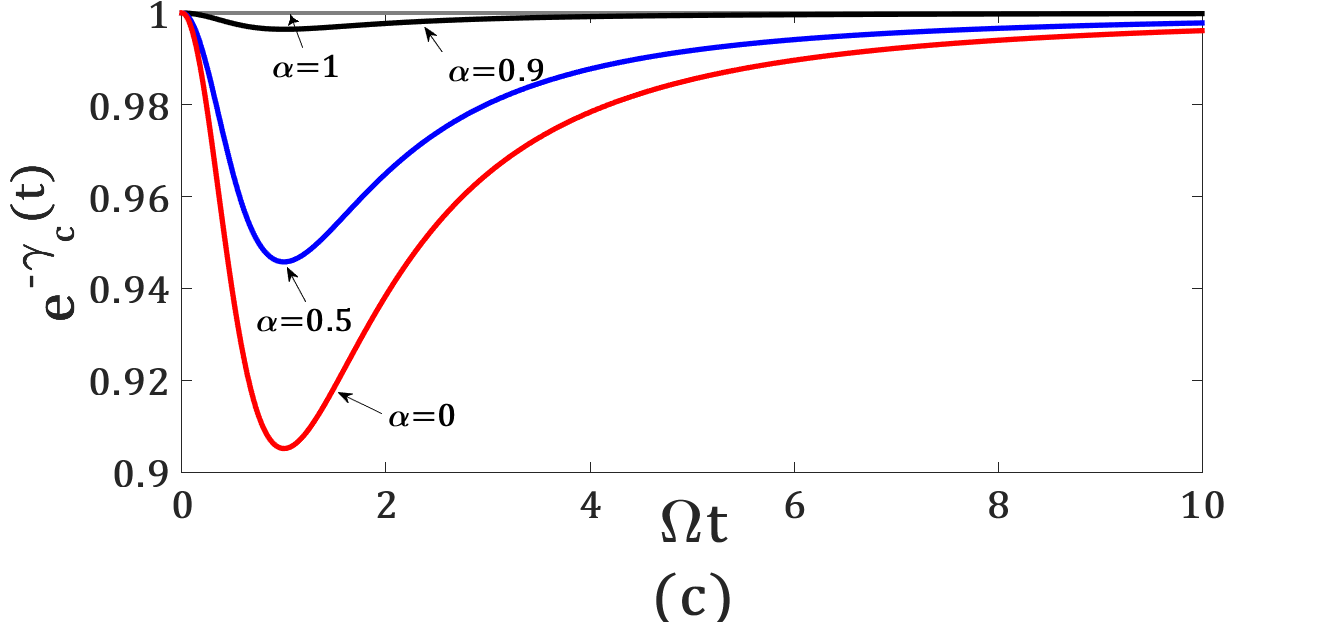}}
\end{center}
 \caption{Time dependence of   exponential of the decoherence due to initial correlations $e^{-\gamma_c(t)}$ 
 for different
 values of $\alpha$  
 with $\beta\omega_0=1$ and the initial condition $\langle\sigma_z\rangle=0, \Omega\beta=1$
 in all the  (a) subohmic  s=0.2 (b) ohmic  s=1 and (c) superohmic  s=2 cases.}
 \label{gamma_c}
\end{figure*}
Next, in order to understand the effect of parameter $\alpha$ on decoherence dynamics, we define 
bath spectral density function $J(\omega)=\sum_k4|g_k|^2\delta(\omega-\omega_k)$. It is convienent 
to describe $J(\omega)$ phenomenologically by assuming the power law form with certian frequency cut-off. Therefore
we write $J(\omega)=\lambda_s (\omega/\Omega)^s \Omega e^{-\omega/\Omega}$ where $\lambda_s$ is the dimensionless
coupling constant and $\Omega$ is the cutt-off frequency. The values $s$ determine the nature of the bath, 
if $s=1$, we call it as ohmic bath while $s<1$ and $s>1$ are called as sub-ohmic and super-ohmic baths respectively.

 Figure \ref{Fig1}
 shows the variation of  total decoherence 
$e^{-\gamma (t)}$ with respect to $\Omega t$ for different values of $\alpha$ in the 
 subohmic $s<1$ , ohmic $s=1$ and superohmic $s=2$ regimes, where $\gamma(t) = \gamma_1(t) + \gamma_c(t)$.
We see from figure \ref{Fig1}(a), that for $\alpha=0$  we observe  strong  oscillations of $e^{-\gamma(t)}$. However 
as $\alpha$ increases from 0 to 1  , the oscillations freeze out. 
The oscillations observed in $e^{-\gamma(t)}$ are due the initial correlations 
in the subohmic regime
 as can be seen from figure \ref{gamma_c}(a). 
The period of oscillations increases from finite to infinite value, thus freeze out at 
the value of $\alpha=1$. 
These features are due to the fact that it takes a longer
time for the system to complete one Hilbert space oscillation resulting in extremely slow dynamics
near the boundary $\alpha=1$ on which the dynamics completely freezes. 



Next, we turn to the ohmic case $s=1$. In this case, using the explicit form of $J(\omega)$, 
we can write the explicit form of decoherence functions in closed form as \cite{mozorov}
 \begin{eqnarray}
  \gamma_1(t)&=& (1-\alpha^2)\left[ \frac{\lambda_1}{2} \ln(1+\Omega^2 t^2) \right.\nonumber \\
 &&\left. \!\!\!\!+2\lambda_1[\ln\Gamma(1+1/\Omega\beta)-\frac{1}{2}\ln\vert\Gamma(1+1/\Omega\beta+it/\beta)\vert^2 \right] \nonumber \\
 \Phi(t)&=&\lambda_1\tan^{-1}(\Omega t)
 \end{eqnarray}
Figures \ref{Fig1}(b) and \ref{gamma_c}(b) show the variation of $e^{-\gamma(t)}$ and  $e^{-\gamma_c(t)}$ with respect to $\Omega t$ for different values of 
$\alpha$. It can be seen from the plot as the hermiticity parameter increases from 0 to 1, the slowing down of 
decoherence is observed. It is noted the sudden transition at $\alpha=1$ with no decoherence at all. 
This feature can be 
attributed to the frozen dynamics of the system Hamiltonian at $\alpha=1$. 

For superohmic case, no oscillatory behaviour is found unlike sub ohmic case see figures \ref{Fig1}(c) and \ref{gamma_c}(c).
This is in complete contrast with the sub ohmic case where 
the oscillation time period becomes infinite.
Although,
in both super-ohmic and sub-ohmic cases as dynamics kicks off at $t=0$ the bath 
has a rapid correlation dependent effect  on the qubit explaining the initial minima in the graphs,
but in  long time limit the qubit settles in a steady state, completely independent of the initial dynamics for superohmic
case, while as no such steady state is formed for sub-ohmic case. 
Nevertheless, both the dynamics have
the same physical consequences that intial correlations disappear. In other words
it would not matter whether initially the system was prepared independently or a projective 
measurement was made on the thermalized system and bath, both will result in approximately same dynamics { near the boundary of separation of physical and unphysical hamiltonians}.


\section{Conclusions}
\label{3}
In conclusion, we studied a $\mathcal{PT}$-symmetric qubit coupled to a bosonic bath.
Using a cannonical transformation, we mapped the $\mathcal{PT}$-symmetric model to well
known spin-boson model which is a purely dephasing model. 
Using projective measurement on system only in a thermalized state of system plus bath, we arrive at 
correlated initial state with bath state depending on the degrees of freedom of the system.
We show that the decoherence due to these initial correlations are strongly modified in the
sub-ohmic regime. Moreover, it is shown that total decoherence is slowed down with the increase
in the hermiticity parameter $\alpha$. At the transition point that seperates the hermitian and non-hermitian regimes,
the dynamics of the qubit freezes out making the qubit more robust against external perturbations. A similiar dynamics
is also observed in Kibble-Zurek mechanism applied to one dimensional Ising model \cite{Ising}.
We see that the decoherence due to initial correlations in all the subohmic, ohmics and super-ohmic 
cases is suppressed in the physically  relevant regime for $\alpha$ near to 1. This results in approximately the
same  dynamics of an initially correlated  and an uncorrelated
 state.

\acknowledgements
MQL acknowledges useful suggestions by Tim Byrnes. JMB acknowledges the hospitality and support at the department of Physics,
University of Kashmir where most of this work was done.

\appendix
\section{}
\label{apen}
In this appendix, we derive the time evolution of the reduced density matrix $\rho_s(t)$ given in the equation
\ref{reduced}. Since the Hamiltonian $\tilde{H}$ given in equation \ref{sp} is a purely dephasing model, which 
results in no dynamics of the  diagonal terms of the density matrix $\rho_s(t)$.
Observing that 
\begin{align*}
e^{-\beta \tilde{H}}\mid 1 \rangle &= e^{-\beta \mu} e^{-\beta H^+ } \mid 1 \rangle\\
e^{-\beta \tilde{H}}\mid 0 \rangle &= e^{\beta \mu} e^{-\beta H^- } \mid 0 \rangle
\end{align*}
with $H^\pm =\sum\limits_k \omega_k b_k^\dagger b_k \pm  \sum\limits_kE(g_k b_k+g_k^*b_k^\dagger) $ and 
$\mu= \frac{\omega_o}{2}$, 
we can write the $\langle 1|\rho_s(t)|0\rangle = \rho_s^{10}(t) $ as
 \begin{eqnarray}
\rho_s^{10}(t)
 &=& \frac{ab^*}{Z}(|a|^2 e^{-\beta\mu }{\rm Tr}_B[ e^{ 2\hat{\Lambda}(t)  }e^{-\beta H^+}] \nonumber \\
&&~ ~~~+|b|^2 e^{\beta\mu }{\rm Tr}_B[ e^{ 2\hat{\Lambda}(t)  }e^{-\beta H^-}])
\end{eqnarray}
where the partition function $Z$ is given by
\begin{align*}
Z=|a|^2 e^{-\beta\mu }{\rm Tr}_B[e^{-\beta H^+}]+|b|^2 e^{\beta\mu }{\rm Tr}_B[e^{-\beta H^-}]
\end{align*}
Next we define a unitary transformation $O_{\pm}= \exp[\pm \sum_k \frac{E}{\omega_k}(g_k b_k - g_k^{*} b_k^{\dagger})]$ 
such that $H^\pm =O_\pm^{-1}(H_B-\xi) O_\pm$  with $\xi = \sum\limits_k E^2\frac{|g_k^2|}{\omega_k}$ and 
$2\hat{\Lambda}(t) =O_\pm^{-1}[2\hat{\Lambda}(t)\pm i E^2 \Phi(t)] O_\pm $. Therefore, we write

\begin{align*}
&\!\!\!\!{\rm Tr}_B[ e^{ 2\hat{\Lambda}(t)}e^{-\beta H^\pm}] \nonumber \\ 
&={\rm Tr}_B[O_\pm^{-1}e^{2\hat{\Lambda}(t)\pm i E^2\Phi(t)} O_\pm O_\pm^{-1}e^{-\beta H_B+\beta \xi}O_\pm ]\\
&={\rm Tr}_B[O_\pm^{-1}e^{2\hat{\Lambda}(t)\pm i E^2\Phi(t)}e^{-\beta H_B+\beta \xi} O_\pm ]\\
&=e^{\pm i(1-\alpha^2)\Phi(t)}Z_Be^{-\gamma_1(t)}e^{\beta \xi}
\end{align*}
where
$Z_B= {\rm Tr}_B[e^{-\beta H_B}]$.
This makes  partition function $Z$  to be
\begin{align*}
Z=Z_Be^{\beta \xi}(|a|^2 e^{-\beta\mu }+|b|^2 e^{\beta\mu })
\end{align*}
Substituting all  the above results in the  expression  for the off diagonal element  $\rho_s^{10}(t)$ 
we have 
\begin{align*}
\rho_s^{10}(t)&=ab^*\frac{|a|^2 e^{-\beta\frac{\omega_0}{2} }e^{i(1-\alpha^2)\Phi(t)}+|b|^2 e^{\beta\frac{\omega_0}{2} }e^{-i(1-\alpha^2)\Phi(t)}}{|a|^2 
e^{-\beta \frac{\omega_0}{2}}+|b|^2 e^{\beta\frac{\omega_0}{2} }}e^{-\gamma_1(t)}
\end{align*}

\end{document}